\documentclass[twoside]{article}
\usepackage{fleqn,espcrc2}

\usepackage{graphicx}
\usepackage[figuresright]{rotating}

\newcommand{\AmS}{{\protect\the\textfont2
  A\kern-.1667em\lower.5ex\hbox{M}\kern-.125emS}}

\title{Strange Magnetic Moment of The Nucleon from Lattice QCD}

\author{Nilmani Mathur\address{TRIUMF, 4004 Wesbrook Mall,
Vancouver, British Columbia, V6T 2A3\\}$^{,b}$ and 
        Shao-Jing Dong\address{Department of Physics and Astronomy, University of Kentucky, Lexington, KY 40506-0055}\\(Kentucky Field theory Collaboration)}
       
\begin{document}

\begin{abstract}

We calculate the strange magnetic moment of the nucleon on a quenched
$16^3 \times 24$ lattice at $\beta = 6.0$, and with Wilson fermions at
$\kappa$ = 0.148, 0.152, and 0.154. The strange quark contribution
from the
disconnected insertion is estimated stochastically by employing the
$Z_2$ noise method. Using an unbiased subtraction along with the help of
charge conjugation and hermiticity, we reduce the error by a factor of 2 with negligible overhead.
Our result is $G_{M}^{s} = -0.28\pm0.10 \mu_{N}$.
\vspace{1pc}
\end{abstract}

\maketitle

\section{INTRODUCTION}
Recently there has been a considerable interest on the 
strange magnetic moment of the nucleon ($G_{M}^{s}(0)$) as it can provide important information about the internal quark structure of the nucleon \cite{group}. There are numerous theoretical predictions over a wide range in between $-0.75$ to $+0.30 \mu_{N}$. 
Most of them are negative and are in between $-0.45$ to $-0.25$ (for summaries see \cite{sum,group}). 
On the other hand, recent experiments of SAMPLE collaboration
on parity violating electron scattering \cite{sample1,sample2}  suggest that 
the strange magnetic moment of the nucleon could be positive.
Two of their recent results are $G_{M}^{s}(Q^{2} = 0.1 GeV^{2}) = + 0.23 \pm0.37\pm0.15\pm0.19 \mu_{N}$ \cite {sample1} and $ +0.61 \pm 0.17 \pm 0.21 \pm 0.19 \mu_{N} $ \cite{sample2}. However, combined with data from the deuteron, the new result is
closer to zero with a relatively larger error~\cite{bec00}.

Previously, our group studied the strange quark magnetic moment on a quenched
lattice  and obtained $G_{M}^{s}(0) = -0.36\pm0.20 \mu_{N}$ \cite{group} . We extend 
that calculation and improve the results by employing CH-transformation and unbiased 
subtraction method, to be discussed subsequently. Finally, we will 
present our new results on $G_{M}^{s}(0)$.

\section{FORMULATION}
We follow the same lattice formulation of the electromagnetic vector current as in Ref. \cite{enm,group}. The magnetic moment of 
the nucleon is extracted from the ratio of 
the three and two-point correlation function. Two point function is defined as
\begin{eqnarray}
G^{\alpha \beta}_{NN}(t,\vec{p})&=&\nonumber\\
&&\hspace{-0.5in}\sum_{\vec{x}} e^{-i \vec{p}\cdot \vec{x}}
\langle O\mid{\hbox{T}}\left(\chi^{\alpha}(x)\bar{\chi}^{\beta}(0)\right)
\mid O\rangle,
\end{eqnarray}
and the three point function is 
\begin{eqnarray}
&&G^{\alpha \beta}_{PJ_{\mu}P}(t_{f},t,\vec{p},\vec{p^{\prime}})\,\,=\,\, \sum_{\vec{x_{f}},\vec{x}} e^{-i \vec{p}\cdot \vec{x_{f}} + i \vec{q}\cdot \vec{x}}
\nonumber\\
&&\hspace*{0.3in}
\langle
0\mid{\hbox{T}}\left(\chi^{\alpha}(x_{f})\,J_{\mu}(x)\,\bar{\chi}^{\beta}(0)
\right)\mid 0\rangle.
%
\end{eqnarray}
Here, $J_{\mu}$ is the point-spilt electromagnetic vector current as defined in Ref. \cite{enm}.
Finally, taking the ratio \cite{enm,group},
\begin{eqnarray}
\hspace*{0.3in}{\Gamma_{i}^{\beta\alpha}
G^{\alpha\beta}_{PJ_{j}P}(t_{f},\vec{0}; t,\vec{q}) \over G^{\alpha\alpha}_{PP}(t_{f},\vec{0})}\,
{G^{\alpha\alpha}_{PP}(t,\vec{0}) \over G^{\alpha\alpha}_{PP}(t,\vec{q})}
&&\nonumber\\
&&\nonumber\\
&&\hspace*{-1.2in}
\longrightarrow {\varepsilon_{ijk}q_{k}\over {E_{q} + m}}\, G^{L}_{M}(q^{2}),
\end{eqnarray}
where $G^{L}_{M}(q^{2})$ is the magnetic moment of the nucleon on lattice at four-momentum transfer square $q^{2}$.

The three point function incorporates both connected insertion (CI) and disconnected insertion (DI). CI arises due to valence and connected sea quarks and DI originates from the vacuum polarization of the disconnected sea quarks \cite{liu}. For the nucleon magnetic moment, strange quarks current, $\bar{s}\gamma_{\mu}s$,  contributes only through the DI (Fig. 1), and so, we will concentrate only on DI contribution of Eq. (3).

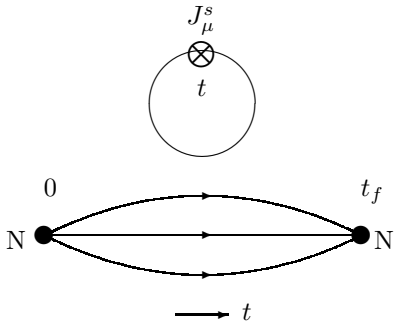
\begin{figure}[h]
\vspace*{-0.6in}
\[
\hspace*{-0.6in}
\setlength{\unitlength}{0.01pt}
\begin{picture}(45000,20000)
\put(8000, 8000){\circle*{700}}
\put(8000, 8000){\line(1,0){12000}}
\put(14000, 8000){\vector(1,0){400}}
\put(20000, 8000){\circle*{700}}
\qbezier(8000, 8000)(14000, 5000)(20000, 8000)
\put(14000, 6500){\vector(1,0){400}}
\qbezier(8000, 8000)(14000,11000)(20000, 8000)
\put(14000, 9500){\vector(1,0){400}}
\put(14000,13000){\circle{4000}}
\put(13400,14600){{\bf $\bigotimes$}}
\put(13400,16000){{\bf $J_{\mu}^{s}$}}
\put(13000, 5000){\vector(1,0){2000}}
\put(13800,13300){$t$}
\put(15500, 4800){$t$}
\put(8000, 9500){$0$}
\put(20000, 9500){$t_{f}$}
\put(6580,7500){N}
\put(20500,7500){N}
%
\end{picture}
\]
\vspace*{-1.2in}
\caption{Disconnected insertion diagrams to show  the strange quark contribution to the nucleon magnetic moment.}
\end{figure}

\vspace*{-0.2in}

\section{LATTICE CALCULATION}         
In general, DI is very difficult to calculate as it involves loop effect 
from the self-contraction of the sea quarks of the current. 
As in the previous DI calculations \cite{di,di1,spin}, for Eq. (3), 
the current insertion time ($t$) from the nucleon source to its sink is 
summed over, and also the average over three spatial directions is considered. 
With this average, this ratio R(3pt,2pt)(Eq. (3)) becomes proportional to 
constant + $t_{f}[|\vec{q}|/E_{q} + m] G_{M}^{DI}(q^{2})$ \cite{group}.
Therefore, by extracting the slope of R(3pt,2pt) {\it vs} $t_{f}$ plot, one can evaluate $G_{M}^{DI}(q^{2})$. 
As shown in Fig. 1, the three point function of 
DI  consists of two parts : nucleon propagator (two-point function) 
and a loop contribution. Similar to other DI calculations \cite{di,di1,spin}, 
to evaluate the trace for the quark loop, we use $Z_{2}$ noise 
trace estimator with a stochastic algorithm. 
It has been shown earlier that by employing 
CH-transformation and unbiased subtraction method one can reduce 
the error in DI substantially \cite{di1,spin}. Similar error reduction procedure is 
adopted here. By CH-transformation we mean the charge conjugation (C) 
and Euclidean hermiticity (H). Under charge conjugation, 
the fermion matrix ($M$) transforms as $M^{-1}(x,y,U)  
=   C^{-1} {\tilde{M}}^{-1}(y,x,U^{*}) C$. Along with this,  
using hermiticity property of 
the fermion matrix ($\gamma_{5}M^{-1}(x,x+a_{\mu})\gamma_{5} = M^{-1\dagger}(x+a_{\mu},x)$), one can show that under CH-transformation, 
\begin{eqnarray}
M^{-1}(x,y,U) = C^{-1}\gamma_{5}{M^{-1}}^{*}(x,y,U^{*})\gamma_{5}C.
\end{eqnarray}
Using this CH-transformation, it is possible to show that the three point function is either real or imaginary, and for the vector current it is real.
Using only Euclidean hermiticity, one can also show that the loop part is imaginary, and thus, after Fourier transformation it becomes real. Therefore, to get the three point function one needs to  multiply only real part (real with Fourier transformation) of loop with the real part (real with Fourier transformation) of the two-point function. Thus, one can avoid  unnecessary $imaginary \times  imaginary$ multiplication, and thereby can eliminate unwanted noises. 

Along with this CH-transformation, as in  angular momentum calculation \cite{spin}, we also adopt an unbiased subtraction method (for details discussion about unbiased subtraction, see Ref. \cite{di1,spin}).  The trace of an inverse matrix $A^{-1}$   can be estimated as :
\begin{eqnarray}
\hbox{Tr} A^{-1} = E {\left[\left< \eta^{\dagger} \left( A^{-1} 
- \sum^{P}_{p=1} \lambda_{p} O^{(P)} \right) \eta \right> \right]},
\end{eqnarray}
where $\eta$s are $Z_{2}$ noise vectors,
$O^{P}$s are a set of traceless matrices,
and $\lambda_{p}$s are variational coefficients which can be tuned to get the minimum 
variance in trace estimation. It was found earlier \cite{di1} that    a set of traceless matrices obtained   from the hopping parameter expansion of the propagaptor 
\begin{eqnarray*}
M^{-1} = I + \kappa D + \kappa^{2} D^{2} + \kappa^{3} D^{3} + \cdots \cdots 
\end{eqnarray*}
can offset the off-diagonal contribution to the variance of the trace, 
and thus they can be used as a possible choice of $O^{(P)}$ matrices. 
For our calculation we use traceless matrices up to order $D^{2}$.

Numerical calculation is done on a quenched $16^{3} \times 24$ lattice at $\beta = 6.0$ with Wilson fermions, and for 100 configurations. Results are obtained for relatively light quarks with $\kappa = 0.148, 0.152,$ and 0.154. For the stochastic trace estimation,  300 noises per configuration is used. 

\newpage
\begin{figure}[h]
\includegraphics{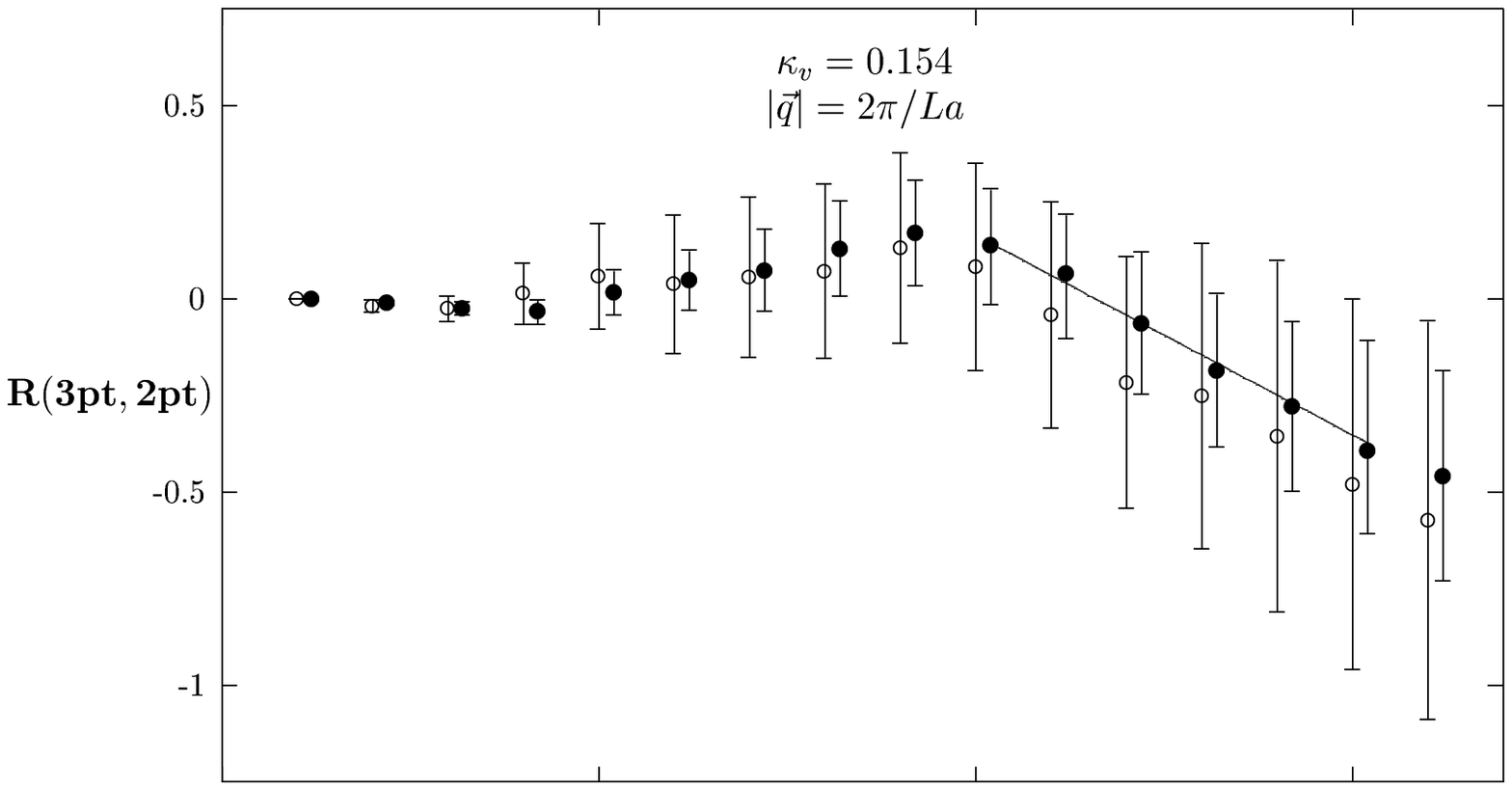}
\end{figure}
\vskip 1.0in
\begin{figure}[h]
\includegraphics{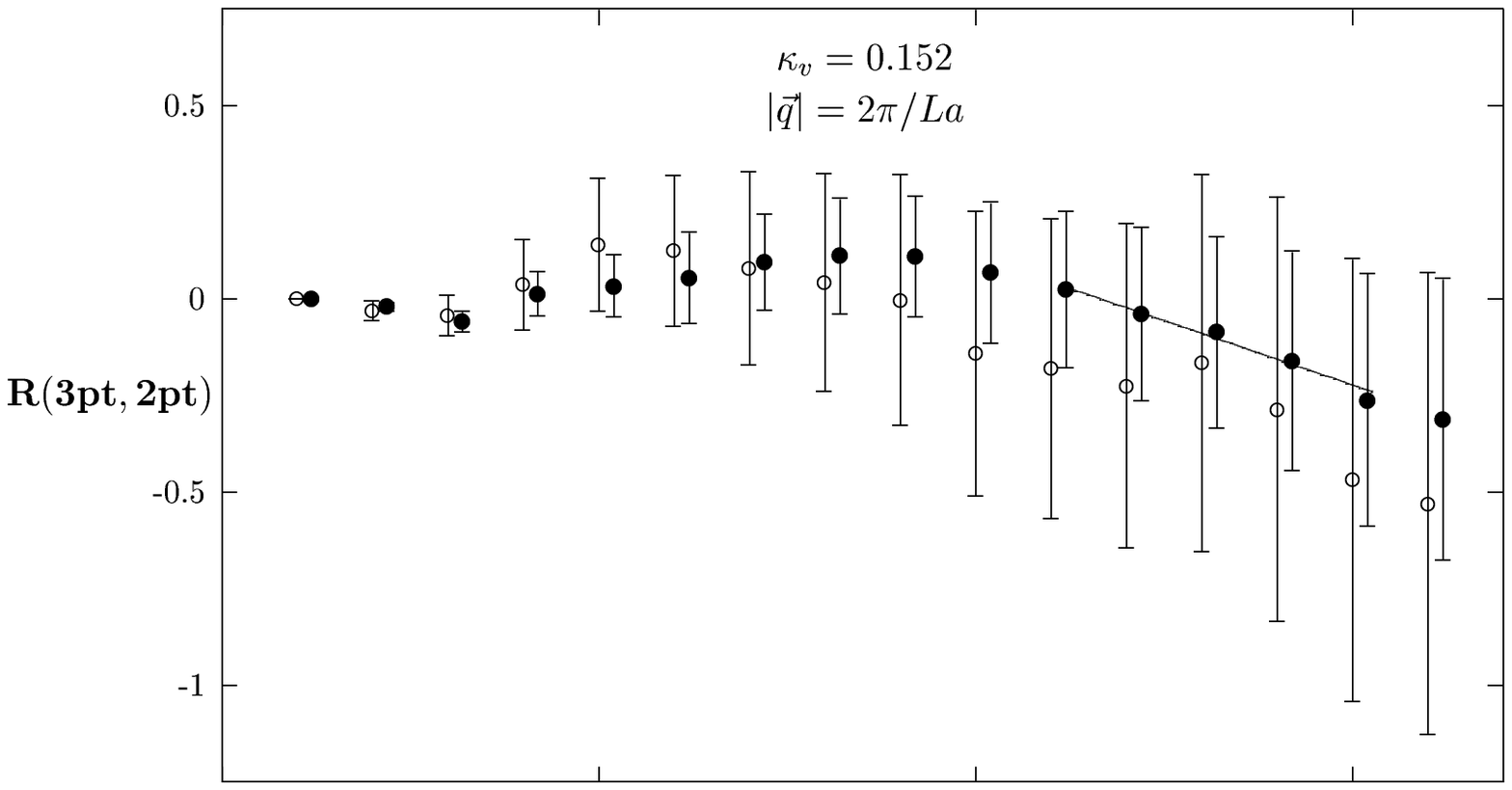}
\end{figure}
\vskip 1.0in
\begin{figure}[h]
\includegraphics{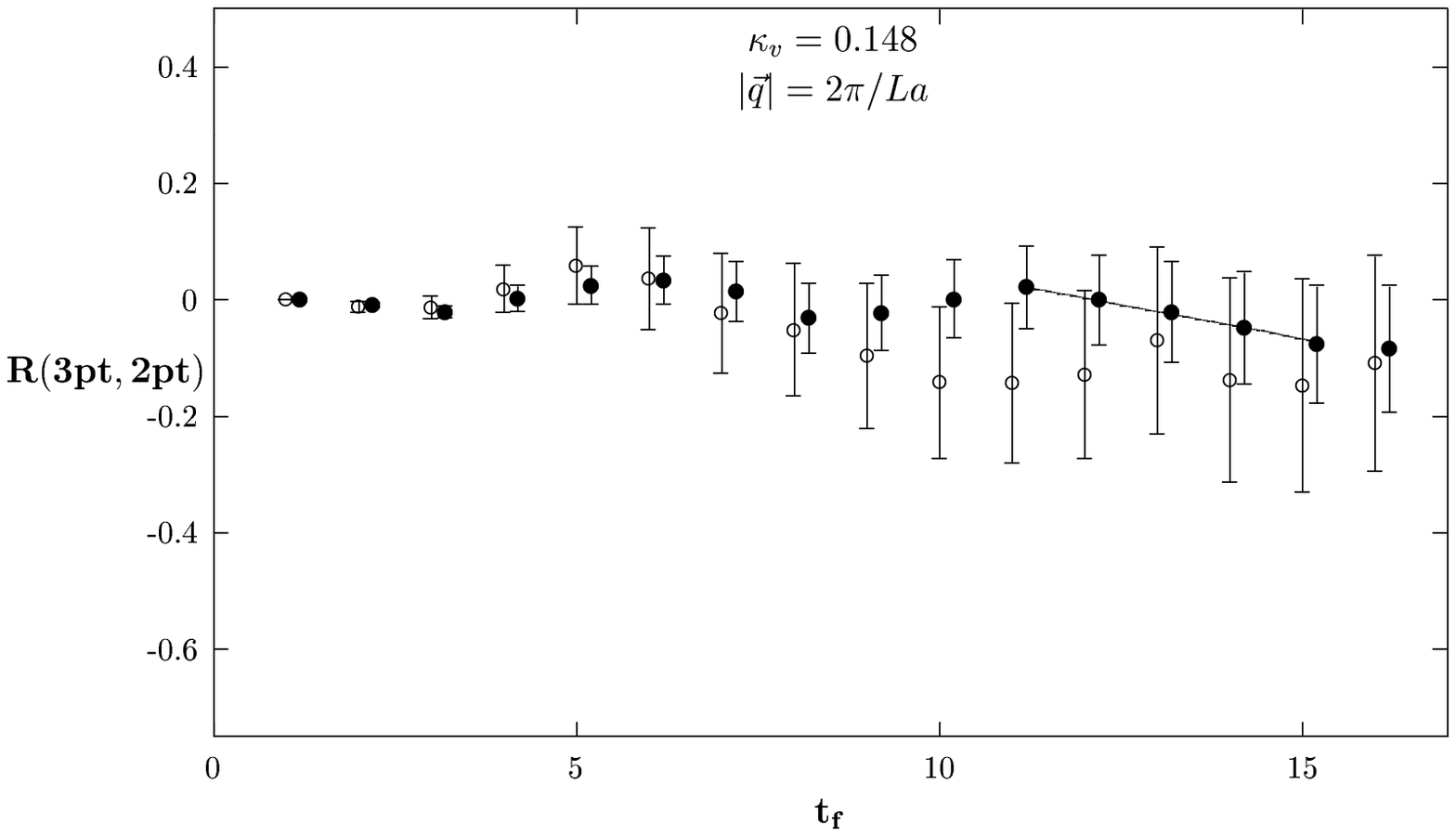}
\vskip 2.0in
\caption{The summed ratios (R(3pt,2pt)) of Eq. (3) are plotted for different final time-slices. Results with and without subtractions (along with CH-transformation) are represented by solid and open circles, respectively. Ratios with subtractions are shifted slightly towards right.}
\end{figure}

\begin{figure}[h]
\vspace*{-0.8in}
\includegraphics{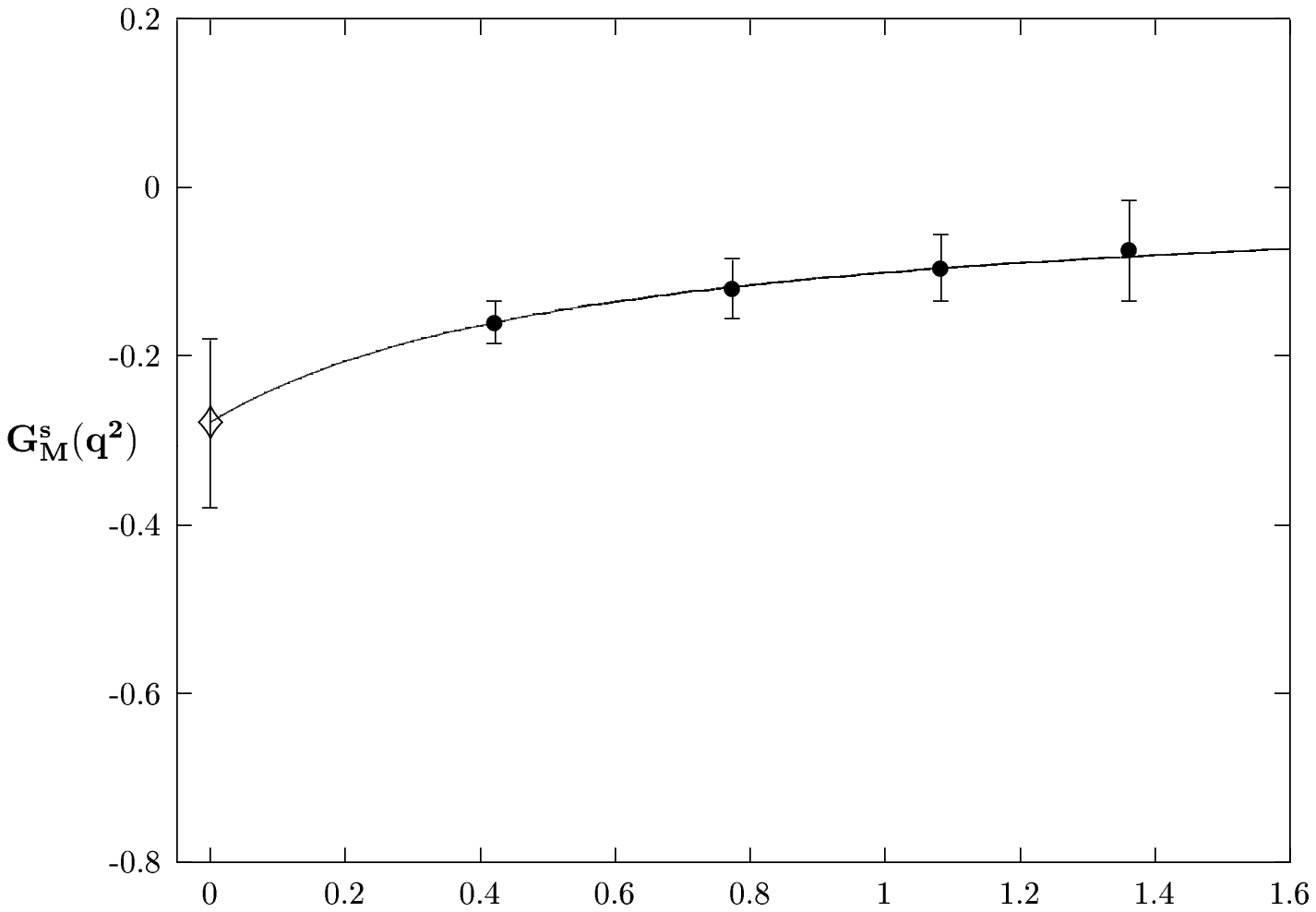}
\vskip -0.4in
\end{figure}
\begin{figure}
\includegraphics{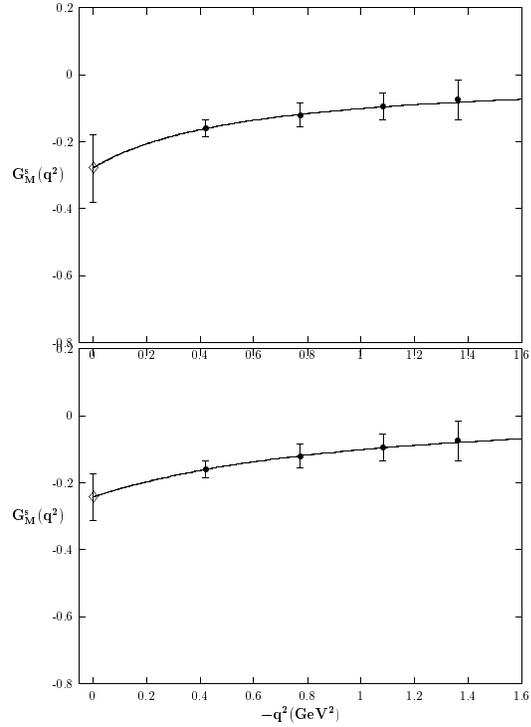}
\vskip 4.8in
\caption{Monopole (top) and dipole (bottom) fitting for the strangeness magnetic form factor $G_{M}^{s}(q^{2})$ of proton. Fitted $G_{M}^{s}(0)$ values are shown by $\diamond$.}
\end{figure}

\section{RESULTS}
As mentioned earlier, the strangeness content of the nucleon magnetic moment  , $G_{M}^{s}(0)$,
appears only in the DI. The numerical procedure for 
extracting $G_{M}^{s}(q^{2})$ has been given in detail in Ref. \cite{group}. 
  First, we fix the sea quark mass (loop mass) at the strange 
quark mass (at $\kappa = 0.154$), and then,  for various valence 
quark masses ($\kappa = 0.148, 0.152$, and $0.154$),  we calculate slopes 
for the summed ratios, R(3pt,2pt), with respect to various final time slices. 
In Fig. (2), we plot these summed ratios as a function of time slices   
(at $|\vec{q}| = 2\pi/L$). As mentioned earlier, the slope of this 
plot provides 
the DI contribution of the magnetic moment, $G^{s}_{M}(q^{2})$. One can easily notice from these plots that we reduce the error-bar of previous calculation \cite{group} 
(ratio with open circle) substantially (by a factor of $\sim 2$). These ratios are calculated at various $q^{2}$, and finally, to obtain  the physical 
$G_{M}^{s}(q^{2})$,   we extrapolate  the valence quark mass to 
the chiral limit with a form $A + B \sqrt{\bar{m}+m_{s}}$, 
where $\bar{m}$ is the average $u$ and $d$ quark mass and $m_{s}$ is 
the strange quark mass, respectively.  This form of 
$m_{s}$-dependence is chosen as $G_{M}^{s}$ is proportional to 
the kaon mass, $m_{k}$. Finally,  to obtain the strange quark contribution to 
the magnetic moment, $G^{s}_{M}(0)$, we fit different $G_{M}^{s}(q^{2})$ 
values with monopole and dipole forms, as shown in  Fig. (3). For 
all these fittings (Fig. 2 and 3), required correlations 
among different quantities are taken into account 
by covariant matrices, and all errors are estimated
from jackknifing the fitting procedure. 

From the monopole fit $G^{s}_{M}(q^{2})$,  
we extract $G^{s}_{M}(0) = -0.28 \pm 0.10 \mu_{N}$, with the monopole mass, $m_{M}/m_{N} = 0.80 \pm 0.09$. It should be recalled that 
the previous result (before subtraction) for $G^{s}_{M}(0)$ was 
$-0.36 \pm 0.20 \mu_{N}$ \cite{group}, {\it i.e.,} we have reduced error by almost
 a factor of 2. Another fit of  $G^{s}_{M}(q^{2})$ with dipole form yields 
$G^{s}_{M}(0) = -0.24 \pm 0.08 \mu_{N},$ and the corresponding dipole mass is
$m_{D}/m_{N} = 1.44 \pm 0.16$. As argued in Ref. \cite{group,di1,spin}, 
we choose the monopole form over dipole for DI. Some other useful quantities,
e.g., the total sea quark contribution, $G_{M}^{sea}(0)$, are also calculated. 
Taking account CI results from Ref.\cite{group}, we also evaluate 
the proton and neutron magnetic moments. Since the total sea quark contribution, $G_{M}^{sea}(0)$, is small, these moments are almost same as in Ref.\cite{group}.
In Table 1, we summarize our result for various quantities. A comparison is also made to show the error reduction. 

\begin{table}[h]
\caption{Nucleon magnetic moment (in $\mu_{N}$) before and after error reduction.}
\begin{tabular}{ccc}
\hline
\hline
&Before \cite{group}            & After \\
\hline
$G^{s}_{M}(0)_{mono}$&$-0.36(20)$& $-0.28(10)$\\
$G^{s}_{M}(0)_{dipo}$& $ -0.27(12)$&$ -0.24(8)$\\
$G^{u(d)}_{M}(0)_{mono}$&$-0.65(30)$& $-0.60(16)$\\
$G^{u(d)}_{M}(0)_{dipo}$& $ -0.45(16)$&$ -0.40(12)$\\
$G^{sea}_{M}(0)$&$-0.097(37)$&$-0.106(28)$\\
$G_{M}(0)_{proton} \simeq$&$ 2.53(10)$&$2.52(8)$\\
$G_{M}(0)_{neutron}\simeq$&$ -1.71(12)$&$-1.71(9)$\\
${G_{M}(0)_{pro}\over G_{M}(0)_{neu}}\simeq$&$-0.68(7)$&$-0.68(5)$\\
&&\\
\hline
\hline
\end{tabular}
\end{table}
\section{CONCLUSIONS}
We reduce the error of strange quark magnetic moment, as estimated previously \cite{group}, by a factor of 2. Strangeness magnetic moment of the nucleon is 
still negative and is $-0.28\pm 0.10 \mu_{N}$. Once more, we prove that the error reduction procedure, by implementing CH-transformation and unbiased subtraction method, works well  in reducing error in the disconnected insertion. Finally, we like to mention that the finding of this work is in disagreement with the results of the next talk \cite{walter}.


\section{ACKNOWLEDGEMENT}This work is partially supported by the Natural Sciences and Engineering Research Council of Canada, DOE Grant No.DE-FG05-84ER40154, and NSF grant No.9722073.


\begin{thebibliography}{9}
\bibitem{group}S. J. Dong, K. F. Liu and A. Williams, Phys. Rev. {\bf D58},074504 (1998).
\bibitem{sum}D. B. Leinweber, Phys. Rev. {\bf D53}, 5115 (1996); C. V. Christov {\it et al.}, Prog. Part. Nucl Phys. {\bf 37}, 1 (1996).
\bibitem{sample1} B. Mueller {\it et al.}, Phys. Rev. Lett. {\bf 78}, 3824 (1997).
\bibitem{sample2}D. Spayde {\it et al.}, Phys. Rev. Lett. {\bf 84}, 
1106 (2000).
\bibitem{bec00}
D. H. Beck, talk at
``Strangeness 2000'', Berkeley, CA, July 20--25, 2000.
\bibitem{enm}W. Wilcox, T. Drapper, and K. F. Liu,  Phys. Rev. {\bf D46}, 1109 (1992). 
\bibitem{liu}K. F. Liu,  Phys. Rev. {\bf D46}, 074501 (2000). 

\bibitem{di}S.J. Dong, J,-F. Lagae, and K. F. Liu, Phys. Rev. Lett. {\bf 75}, 2096 (1995); Phys. Rev. {\bf D54}, 5496 (1996). 
\bibitem{di1}C. Thron, S. J. Dong, K.F. Liu, and H. P. Ying, Phys. Rev. {\bf D57}, 1642 (1998). 
\bibitem{spin}N. Mathur, S. J. Dong, K. F. Liu, L. Mankiewicz, and N. C. Mukhopadhyay, Phys. Rev. {\bf D62}, 114504 (2000).

\bibitem{walter}W. Wilcox, these proceedings.
\end{thebibliography}
\end{document}